\documentclass[letterpaper,twocolumn,english,prc,aps,showpacs,superscriptaddress]{revtex4}

\usepackage{ae,aecompl}
\usepackage[T1]{fontenc}
\usepackage[latin9]{inputenc}
\usepackage{babel}
\usepackage{units}
\usepackage{amsmath}
\usepackage{amssymb}
\usepackage{graphicx}
\usepackage{float}
\usepackage{subfigure}
\usepackage{tikz}
\usepackage{multirow}
\usepackage{ulem}

\makeatletter

\DeclareTextFontCommand{\helvetica}{\fontfamily{phv}\selectfont}

\makeatother

\begin{document}


\title{Study and validation of a new ``3D Calorimetry'' \\ of hot nuclei with the HIPSE event generator}

\author{E. Vient}
\thanks{vient@lpccaen.in2p3.fr; http://caeinfo.in2p3.fr/}
\affiliation{Normandie Univ, ENSICAEN, UNICAEN, CNRS/IN2P3, LPC Caen, F-14000 Caen, France}

\author{L. Manduci}
\affiliation{\'Ecole des Applications Militaires de l'\'Energie Atomique, B.P. 19, F-50115 Cherbourg, France}
\affiliation{Normandie Univ, ENSICAEN, UNICAEN, CNRS/IN2P3, LPC Caen, F-14000 Caen, France}

\author{E. Legou\'ee}
\affiliation{Normandie Univ, ENSICAEN, UNICAEN, CNRS/IN2P3, LPC Caen, F-14000 Caen, France}

\author{L. Augey}
\affiliation{Normandie Univ, ENSICAEN, UNICAEN, CNRS/IN2P3, LPC Caen, F-14000 Caen, France}

\author{E. Bonnet}
\affiliation{SUBATECH UMR 6457, IMT Atlantique, Universit\'e de Nantes, CNRS-IN2P3, 44300 Nantes, France}

\author{B. Borderie}
\affiliation{Institut de Physique Nucl\'eaire, CNRS/IN2P3, Univ. Paris-Sud, Universit\'e Paris-Saclay, F-91406 Orsay cedex, France}

\author{R. Bougault}
\affiliation{Normandie Univ, ENSICAEN, UNICAEN, CNRS/IN2P3, LPC Caen, F-14000 Caen, France}

\author{A. Chbihi}
\affiliation{Grand Acc\'el\'erateur National d'Ions Lourds (GANIL), CEA/DRF-CNRS/IN2P3, Bvd. Henri Becquerel, 14076 Caen, France}

\author{D. Dell'Aquila}
\affiliation{Institut de Physique Nucl\'eaire, CNRS/IN2P3, Univ. Paris-Sud, Universit\'e Paris-Saclay, F-91406 Orsay cedex, France}
\affiliation{Dipartimento di Fisica 'E. Pancini' and Sezione INFN, Universit\'a di Napoli 'Federico II', I-80126 Napoli, Italy}

\author{Q. Fable}
\affiliation{Grand Acc\'el\'erateur National d'Ions Lourds (GANIL), CEA/DRF-CNRS/IN2P3, Bvd. Henri Becquerel, 14076 Caen, France}

\author{L. Francalanza}
\affiliation{Dipartimento di Fisica 'E. Pancini' and Sezione INFN, Universit\'a di Napoli 'Federico II', I-80126 Napoli, Italy}

\author{J.D. Frankland}
\affiliation{Grand Acc\'el\'erateur National d'Ions Lourds (GANIL), CEA/DRF-CNRS/IN2P3, Bvd. Henri Becquerel, 14076 Caen, France}

\author{E. Galichet}
\affiliation{Institut de Physique Nucl\'eaire, CNRS/IN2P3, Univ. Paris-Sud, Universit\'e Paris-Saclay, F-91406 Orsay cedex, France}
\affiliation{Conservatoire National des Arts et M\'etiers, F-75141 Paris Cedex 03, France}

\author{D. Gruyer}
\affiliation{Normandie Univ, ENSICAEN, UNICAEN, CNRS/IN2P3, LPC Caen, F-14000 Caen, France}
\affiliation{Sezione INFN di Firenze, Via G. Sansone 1, I-50019 Sesto Fiorentino, Italy}

\author{D. Guinet}
\affiliation{IPNL/IN2P3 et Universit\'e de Lyon/Universit\'e Claude Bernard Lyon1, 43 Bd du 11 novembre 1918 F69622 Villeurbanne Cedex, France}

\author{M. Henri}
\affiliation{Normandie Univ, ENSICAEN, UNICAEN, CNRS/IN2P3, LPC Caen, F-14000 Caen, France}

\author{M. La Commara}
\affiliation{Dipartimento di Fisica 'E. Pancini' and Sezione INFN, Universit\'a di Napoli 'Federico II', I-80126 Napoli, Italy}

\author{G. Lehaut}
\affiliation{Normandie Univ, ENSICAEN, UNICAEN, CNRS/IN2P3, LPC Caen, F-14000 Caen, France}

\author{N. Le Neindre}
\affiliation{Normandie Univ, ENSICAEN, UNICAEN, CNRS/IN2P3, LPC Caen, F-14000 Caen, France}

\author{I. Lombardo}
\affiliation{Dipartimento di Fisica 'E. Pancini' and Sezione INFN, Universit\'a di Napoli 'Federico II', I-80126 Napoli, Italy}
\affiliation{INFN - Sezione Catania, via Santa Sofia 64, 95123 Catania, Italy}

\author{O. Lopez}
\affiliation{Normandie Univ, ENSICAEN, UNICAEN, CNRS/IN2P3, LPC Caen, F-14000 Caen, France}

\author{P. Marini}
\affiliation{CEA, DAM, DIF, F-91297 Arpajon, France}

\author{M. P\^arlog}
\affiliation{Normandie Univ, ENSICAEN, UNICAEN, CNRS/IN2P3, LPC Caen, F-14000 Caen, France}
\affiliation{Hulubei National Institute for R$\And$D in Physics and Nuclear Engineering (IFIN-HH), P.O.BOX MG-6, RO-76900 Bucharest-M\`agurele, Romania}

\author{M. F. Rivet}
\thanks{deceased}
\affiliation{Institut de Physique Nucl\'eaire, CNRS/IN2P3, Univ. Paris-Sud, Universit\'e Paris-Saclay, F-91406 Orsay cedex, France}

\author{E. Rosato}
\thanks{deceased}
\affiliation{Dipartimento di Fisica 'E. Pancini' and Sezione INFN, Universit\'a di Napoli 'Federico II', I-80126 Napoli, Italy}
\author{R. Roy}
\affiliation{Laboratoire de Physique Nucl\'eaire, Universit\'e Laval, Qu\'ebec, Canada G1K 7P4}

\author{P. St-Onge}
\affiliation{Laboratoire de Physique Nucl\'eaire, Universit\'e Laval, Qu\'ebec, Canada G1K 7P4}
\affiliation{Grand Acc\'el\'erateur National d'Ions Lourds (GANIL), CEA/DRF-CNRS/IN2P3, Bvd. Henri Becquerel, 14076 Caen, France}

\author{G. Spadaccini}
\affiliation{Dipartimento di Fisica 'E. Pancini' and Sezione INFN, Universit\'a di Napoli 'Federico II', I-80126 Napoli, Italy}

\author{G. Verde}
\affiliation{Institut de Physique Nucl\'eaire, CNRS/IN2P3, Univ. Paris-Sud, Universit\'e Paris-Saclay, F-91406 Orsay cedex, France}
\affiliation{INFN - Sezione Catania, via Santa Sofia 64, 95123 Catania, Italy}

\author{M. Vigilante}
\affiliation{Dipartimento di Fisica 'E. Pancini' and Sezione INFN, Universit\'a di Napoli 'Federico II', I-80126 Napoli, Italy}

\vspace{0.5cm}
\collaboration{INDRA collaboration}
\noaffiliation

\date{\today}%
\begin{abstract}
In nuclear thermodynamics, the determination of the excitation energy of hot nuclei is a fundamental experimental problem.  Instrumental physicists have been trying to solve this problem for several years by building the most exhaustive  $4\pi$ detector arrays and perfecting their calorimetry techniques. In a recent paper, a proposal for a new calorimetry, called ``3D calorimetry'', was  made. It tries to optimize the separation between the particles and fragments emitted by the Quasi-Projectile and the other possible contributions. This can be achieved by determining the experimental probability for a given nucleus of a nuclear reaction to be emitted by the Quasi-Projectile.  It has been developed for the INDRA data. In the present work, we wanted to dissect and validate this new method of characterization of a hot Quasi-Projectile. So we tried to understand and control it completely to determine these limits. Using the HIPSE event generator and a software simulating the functioning of INDRA, we were able to achieve this goal and provide a quantitative estimation of the quality of the QP characterization.
\end{abstract}

\maketitle

\section{\label{sec1}INTRODUCTION}
Heavy-ion collisions in the Fermi energy domain allow the formation of very hot nuclei. So we want to study the evolution of these hot nuclei, as the deposited energy increases. We should be able to observe a ``phase transition'' of these drops of nuclear matter. But we must keep in mind that the collision processes in this energy range are very complex. In this energy region of 10 - 100 A.MeV, there is a competition between the effects of the nucleus mean field and the nucleon-nucleon interaction \cite{Lehaut1, Lopez1}. This leads  to an evolution of reaction mechanisms \cite{Fuchs1}, which gradually develop from complete (incomplete) fusion  or deeply inelastic collisions with a statistical emission towards binary collisions \cite{Steck1, Lott1, Baldwin1, Bougault1} accompagnied by particle emissions at different equilibration degrees,  neck emission \cite{Stuttge1, Toke1, Lukasik1, Bocage1, DiToro1} and pre-equilibrium  \cite{ Germain1, TLefort1}. It is in this context that the nuclear physicists by means of calorimetry \cite{Viola2}, must try to isolate the hot nuclei formed during these processes and to characterize them in the best possible way.  ``3D calorimetry'' is presented in detail in two references \cite{Vient1, Vient2}. The basic idea is that, in a restricted area of the velocity space in the QP frame, we can completely define the spatio-energetic characteristics of the evaporated nuclei by the QP. By comparison with the other areas of the velocity space, the probability for a nucleus produced in the collision to be evaporated by the QP can be determined. Then, event by event,  these probabilities are used to reconstruct the QP, \textit{i.e.} to determine its mass, charge, velocity in the center of mass (c.m.) frame and excitation energy. This calorimetry was applied to the data of Xe+Sn reactions acquired with the INDRA detector array \cite{Pouthas1} at different incident energies. The presented work is devoted to the study and validation of this new experimental calorimetry using the HIPSE event generator \cite{Lacroix1}. In addition, a software simulating as realistically as possible all the detection phases by INDRA \cite{Copinet1, Cussol1} will be used. All the events supplied by HIPSE can be filtered by it. 
 In the first section, the event generator used is qualified by a quantitative comparison with the data obtained by the INDRA collaboration for the system of interest. We want verify that it is realistic and correct enough to be used for calorimetry validation. In the following section, we apply the new calorimetry to experimental data  and filtered events provided by HIPSE. By means of a systematic comparison between both, we will quantitatively verify  the quality of the characterization of the hot QP. In the last section, we conclude about this work.  

\section{The qualification of the HIPSE event generator\label{sec2}}
The event generator, HIPSE (Heavy Ion Phase Space Exploration), is described in detail in the reference \cite{Lacroix1}. 

HIPSE, compared to commonly used generators, is capable of reproducing most of the processes occurring in heavy-ion collisions at intermediate energies. This is indeed a fundamental contribution. It can manage the pre-equilibrium phase, the physics of the participating zone and the phase of de-excitation of the formed hot nuclei, taking into account all the spatio-temporal correlations due to the Coulomb interaction. It thus allows us to go further in the study of the experimental calorimetry.

It is important for our study to check that the event generator is able to  correctly take into account the static and dynamic characteristics of collisions studied. This has already been done in part in the reference \cite{Lacroix1}. In this article, it is clear that HIPSE, for the system Xe + Sn, is capable of correctly reproducing the charge distributions, the correlations between the charge and the parallel velocity of nuclei produced, the mean kinetic energy according to the charge or to the angle of emission in the center of mass. It allows to find at 50 A.MeV the distribution of the flow angle for complete events, for which 80\% of the initial charge and linear momentum have been detected,  as well as the densities of particles per unit of velocity along the beam, for different types of particles, giving a spatial-energetic correlations close to those of the experimental ones. This is fundamental for the analysis we want to perform with the data supplied by this generator.
It is also important to note that the requirements used in HIPSE to describe the formation of the clusters at mid-rapidity or the pre-equilibrium are also applied in the nIPSE nucleon-nucleus collision model \cite{Lacroix2}. It is also capable of reproducing very well the physical characteristics of the particles produced during such a collision, as shown by \cite{Lacroix2}.
To pass all generated events through the INDRA multi-detector, the software simulating the entire operation of this apparatus and described in the reference \cite{Copinet1}, is always used.
The generated events, detected by this virtual detector, will be called filtered events.
\subsection{General characteristics of the collisions\label{ssec2.1}}
\begin{figure}[htbp]\centerline{\includegraphics[width=8.3cm,height=8.81cm]{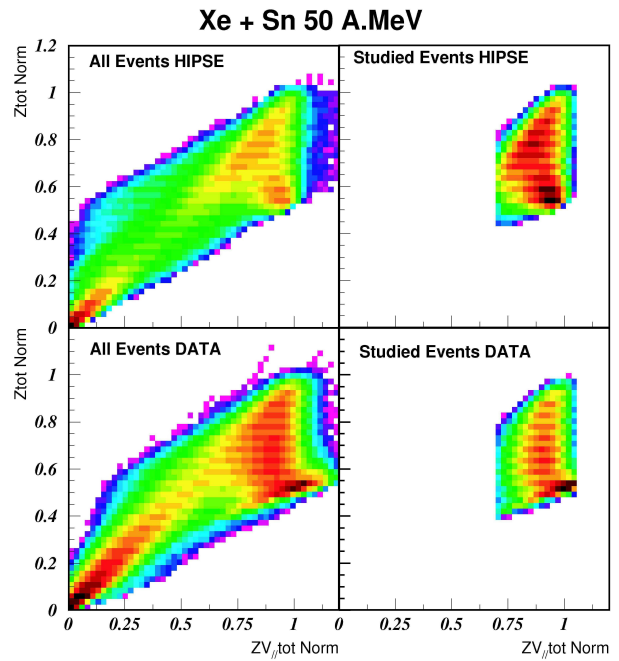}}
\caption{Two-dimensional graphs giving the total detected charge normalized to the initial total charge as a function of the detected pseudo-momentum normalized to the total initial pseudo-momentum (the z scale is logarithmic). This is made for the data and for the generator HIPSE, in two cases: with all the events seen by INDRA and with the events said ``complete'' at the front of the center of mass.\label{fig1}}
\end{figure}

\begin{figure}[htbp]\centerline{\includegraphics[width=6.8cm,height=12.39cm]{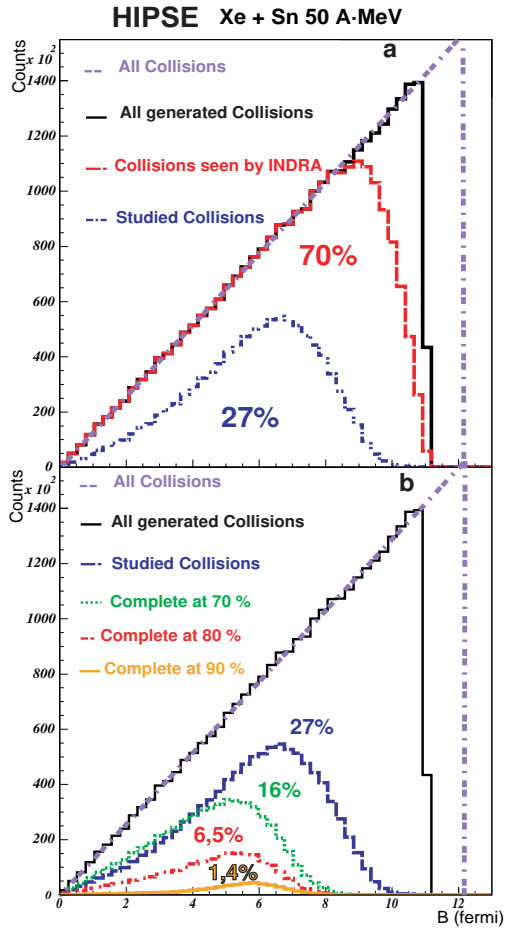}}
\caption {Distributions of impact parameters generated by HIPSE for different criteria of  event completeness. \textbf{a)} For all the generated events, for events with at least one detected particle and for studied events, \textit{i.e.} a completeness at the front of the c.m.. \textbf{b)} For all generated events, for studied events, for events with different total completeness.\label{fig2}}
\end{figure}

 First, we will see how the selections of the events used to perform calorimetry of nuclei, can be used equally for data and  HIPSE. With the help of HIPSE, we also want to estimate the cross sections of detected events and the impact parameters actually studied.
In order to verify the quality of the reaction detection \cite{Steck2, Vient1}, the figure \ref{fig1} shows, for filtered events, the two-dimensional graphs giving the total detected charge normalized to the initial total charge as a function of the detected pseudo-momentum normalized to the total initial pseudo-momentum, on the left for all the events actually detected by INDRA and on the right for the ``studied events'', \textit{i.e.} complete events at the front of the center of mass.
 For the latter, it is the total pseudo-momentum of particles and fragments, located at the front of the c.m., which is normalized  to the initial pseudo-momentum of the projectile. We note a very good qualitative agreement between simulation and data.  
Figure \ref{fig2} shows the distributions of the impact parameter given by HIPSE for various event selections in collisions Xe + Sn at 50 A.MeV. We generated 3 million events with impact parameters between 0 and 11 fermis closed for this study. INDRA detected only 85 \% of these events, \textit{i.e.} about 70 \%  of the total geometrical cross section. Very peripheral collisions are almost invisible for this incident energy. Our criterion of completeness at the front of the center of mass has further reduced our events, as only about 1/3 of the HIPSE events are selected.

It is important to note that this loss is present over the full range of impact parameters as shown again in figures \ref{fig2}-a and \ref{fig4}-a. Various indicative criteria for completeness are presented in figure \ref{fig2}-b. We immediately see that when we ask for completeness not only in the forward hemisphere of the center of mass but in the whole velocity space, the proportion of the events that satisfy the selection decreases sharply because of the very high detection threshold for emissions from the \textbf{Q}uasi-\textbf{T}arget (\textbf{QT}). Finally, only 16\%, 6.5\% and 1.4\% of total geometrical cross section are retained for completeness requirements of 70\%, 80\% and 90\% respectively. These criteria eliminate all the events with an impact parameter greater than 8 fm. The  80\% completeness criterion has been widely used by the INDRA collaboration. A study, performed by N.Marie and al. in reference \cite{Marie2} on the data Xe + Sn at 50 A.MeV, showed that such a selection represents 12.5 \% of the events effectively seen by INDRA. If we do an equivalent calculation with HIPSE, we find a close but not identical result of 9.3 \%.  

Experimentally we do not have direct access to the impact parameter. But, it can be deduced from an experimental observable strongly correlated with it, see for example the references  \cite{Peter1,Marie2,Cavata1, Bowman1, Phair1}. In particular, N.Marie et al \cite{Marie2} used the transverse kinetic energy of the LCPs assuming a maximum impact parameter of 12.2 fm.
\begin{figure}[htpb]\centerline{\includegraphics[width=8.6cm,height=8.45cm]{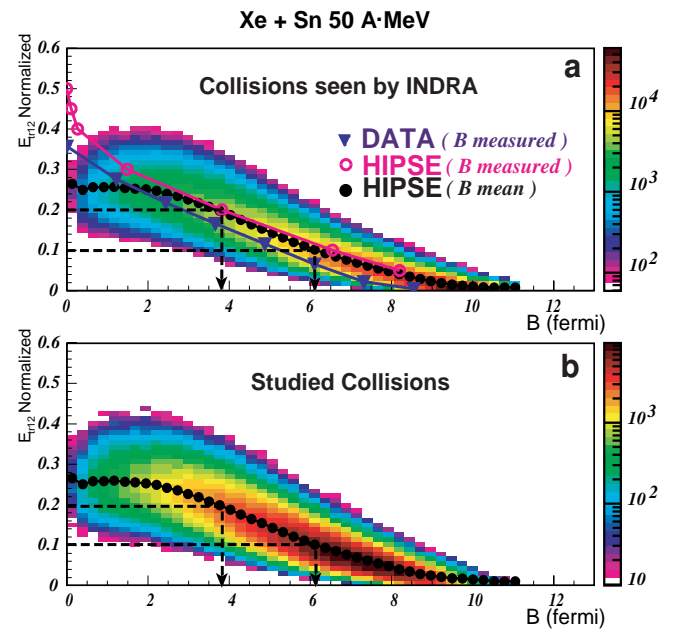}}
\caption{\textbf{a)}  Two-dimensional graph of the normalized transverse kinetic energy of the particles of $Z\leq2$ as a function of the impact parameter, provided by HIPSE for  Xe + Sn at 50 A.MeV, for all the events seen by INDRA. The blue triangles correspond to the average correlation between normalized $E _{tr12}$  and $B$ determined by N.Marie et al in the reference \cite{Marie2}. The pink open circles correspond to the result of the appliying experimental technique used in \cite{Marie2} to the data provided by HIPSE. The full black circles, correspond to the average correlation calculated from the two-dimensional graph. \textbf{b)} For HIPSE only, as a) part, for complete events in the forward hemisphere of the center of mass seen by INDRA. The full black circles correspond to the average correlation calculated from the two-dimensional graph. \label{fig3}}
\end{figure}
 Figure \ref{fig3} shows the correlation of this transverse kinetic energy $E_{tr12}$, normalized to the available energy perpendicular to the direction of the beam in the center of mass, as a function of the impact parameter $B$, for the system $^{129}$Xe + $^{nat}$Sn at 50 A.MeV, as supplied by HIPSE.  We also drew the mean correlation between these two variables (full black  circles). The correlation is shown for all events detected by INDRA in figure \ref{fig3}-a and for those selected by the specific completeness, defined above, at the front of the center of mass in figure \ref{fig3}-b. In figure \ref{fig3}-a, we can see the mean values of this correlation: the blue triangles represent the result of the calculation applied to the data, performed in \cite{Marie2}  and the pink open circles represent the mean values extracted from HIPSE data with the same method. As can be seen, the agreement between the mean correlation (full black circles) and the calculated average (pink open circles) for HIPSE data is good especially between 2 and 8 fermis.
\begin{figure}[htbp]\centerline{\includegraphics[width=6.8cm,height=12.39cm]{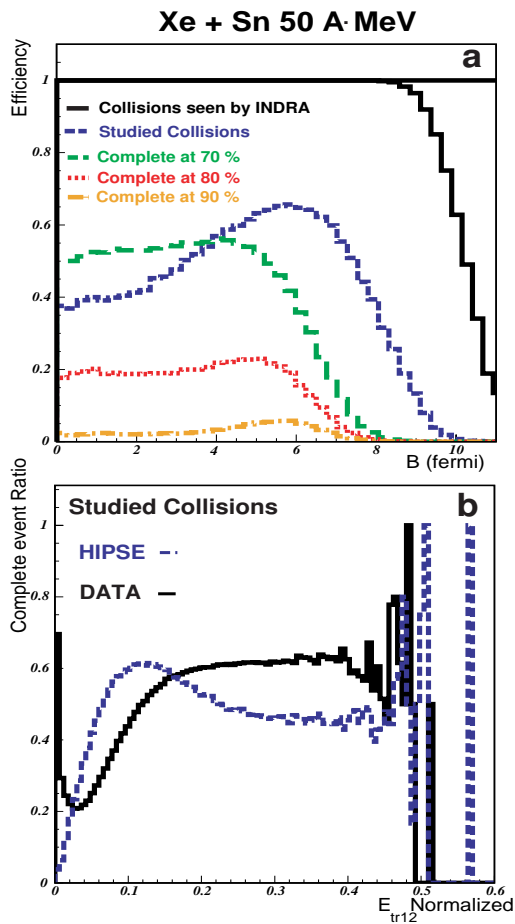}}
\caption{\textbf{a)} Average detection efficiency as a function of the impact parameter for differient completeness criteria obtained from HIPSE filtering for the system Xe + Sn at 50 A.MeV. \textbf{a)} Comparison of the average detection efficiency as a function of the normalized $E_{tr12}$ between HIPSE and data.\label{fig4}}
\end{figure}
 This method is based on the assumption that the efficiency of detecting an event is independent from the impact parameter. Figure \ref{fig4}-a shows that this hypothesis is not true beyond 8.5 fermis closed. In figure \ref{fig3}, it is important to note that the average correlation for the studied events, in HIPSE, is the same that for all the events detected by INDRA. Completeness does not interfere the existing correlation.
On the other hand, figure \ref{fig3} shows a systematic discrepancy between data and HIPSE, even if the evolution of the curve is very similar. To understand this difference, we can refer to  figure \ref{fig5}, where the normalized transverse kinetic energy distributions of LCPs are shown for experimental and simulated events. The number of events studied (blue graph) shows a large difference between data and HIPSE. For data, there is an strong contribution for very low values of normalized $E_{tr12}$, \textit{i.e.} for very peripheral collisions, which is not the case in the simulation. This seems to indicate, for HIPSE, that the Quasi-Projectiles, for these very peripheral collisions, are not correctly detected. In addition, we note that HIPSE gives  maximum values of  transverse kinetic energy slightly higher than the data.

\begin{figure}[htpb]\centerline{\includegraphics[width=8.6cm,height=8.65cm]{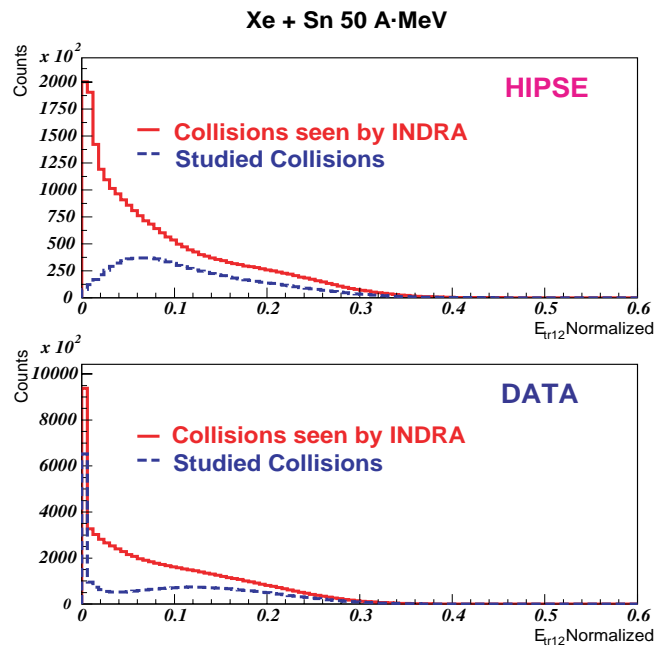}}
\caption{Comparison of the distributions of normalized $E_{tr12}$ for HIPSE and data for all the detected events and ``studied events''.\label{fig5}}
\end{figure}
 The figure \ref{fig4} confirms this finding. In the latter at the top, the efficiency of event detection, supplied by HIPSE, is presented according to the impact parameter for different completeness criteria. At the bottom, in figure \ref{fig4}-b, there are the experimental rates of detection of the ``studied collisions'', \textit{i.e.} complete events in the forward hemisphere of the c.m., as a function of the normalized $E_{tr12}$ for the data and HIPSE. They were obtained by the ratio of the two curves of the figure \ref{fig5}.
It is immediately noticeable that for HIPSE, the detection efficiency as a function of the normalized $E_{tr12}$ obtained for the ``studied events''  is very similar to that obtained as a function of the impact parameter $B$ for these same events, knowing that $E_{tr12}$ increases when the impact parameter $B$ decreases (both curves in blue dotted line). This means that the  efficiency of data experimental detection is better for central collisions, whereas it is less effective for peripheral collisions than for HIPSE (with the exception of the very peripheral collisions).

\begin{figure}[htbp]\centerline{\includegraphics[width=9.6cm,height=8.2cm]{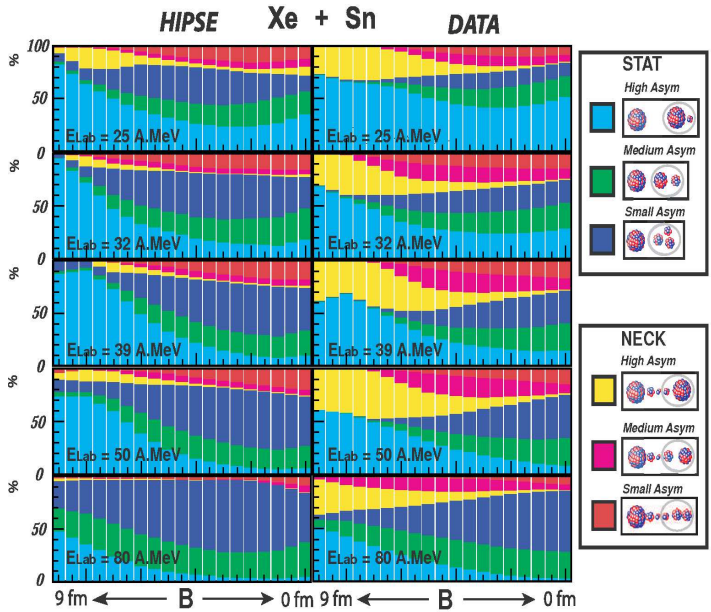}}
\caption{Proportions of the various types of reaction mechanisms according to the ``experimental impact parameter'' (violence of the collision defined by $E_{tr12}$ normalized) for collisions Xe + Sn at 25, 32, 39, 50, 80 A.MeV.}
\label{fig6}
\end{figure}

We also want to study whether HIPSE can correctly reproduce the different reaction mechanisms (\textbf{Statistical} events or events with \textbf{Neck} emission), which have already been shown and discussed in the references \cite{Vient2, Colin1, Normand1}, according to the deduced impact parameter and for collisions Xe + Sn at different incident energies.  To determine the type of reaction mechanism, the same criterion as in reference \cite{Vient2} is used. We observe the presence (or not) of the second heaviest fragment of the forward part of the center of mass, in the forward hemisphere of the QP frame (see figure \ref{fig6}). In this case, it is a \textbf{Statistical} event, corresponding to a statistical emission by the hot nucleus. Otherwise, it is a \textbf{Neck} emission. But we must bear in mind that statistical emission is isotropic. Therefore, we selected only half of the statistical contribution. The other one is necessarily emitted backwards and must therefore be removed from our selection of \textbf{Neck} emission, to obtain a selection corresponding to a ``\textbf{pure Neck} emission''. This is done specifically and only in figure \ref{fig6}.  

At the same time, we studied another aspect: the asymmetry between the two heaviest fragments in the forward part of the center of mass. This is used to obtain the signal of bimodality associated with a liquid-gas phase transition in nuclei \cite{Pichon1, Bonnet2, Bruno1}.  We should sign with this variable a possible transition between evaporation and multifragmentation when the Quasi-Projectile is de-excited.
Figure \ref{fig6} shows the relative proportions of these different mechanisms according to an ``experimental'' impact parameter.  The discrepancy between the data and the HIPSE events can be seen: in particular HIPSE does not produce as much as \textbf{Neck} emissions as experimental data. This could due to an high clusterization rate. The discrepancy appears to be particularly large for peripheral collisions. There is a slight improvement with the centrality of the collision for all the incident energies. It therefore seems necessary to improve the nuclear potential used. In this version of HIPSE, the influence of angular momentum for the clusterization in term of $\ell_{crit}$ has not been optimized.  The discordance is most evident at 80 A.MeV. The participant-spectator aspect becomes too important. 

\subsection{Study of the heaviest fragments at the front of the center of mass\label{ssec2.2}}
Even if the proportions of events do not seem to be well reproduced, we will see that there is nevertheless a good match of physics for all different selections of events. We will first focus  on  studying  the static and dynamic physical characteristics of the two heaviest fragments located at the front of the center of mass. We chose to study these physical quantities according to the violence of the collision using the transverse kinetic energy of LCPs. We have divided its experimental distribution into ten numbered bins of same cross section, allowing for a selection of events according to the bin number associated with the event.
\begin{figure}[htbp]\centerline{\includegraphics[width=8.6cm,height=14.76cm]{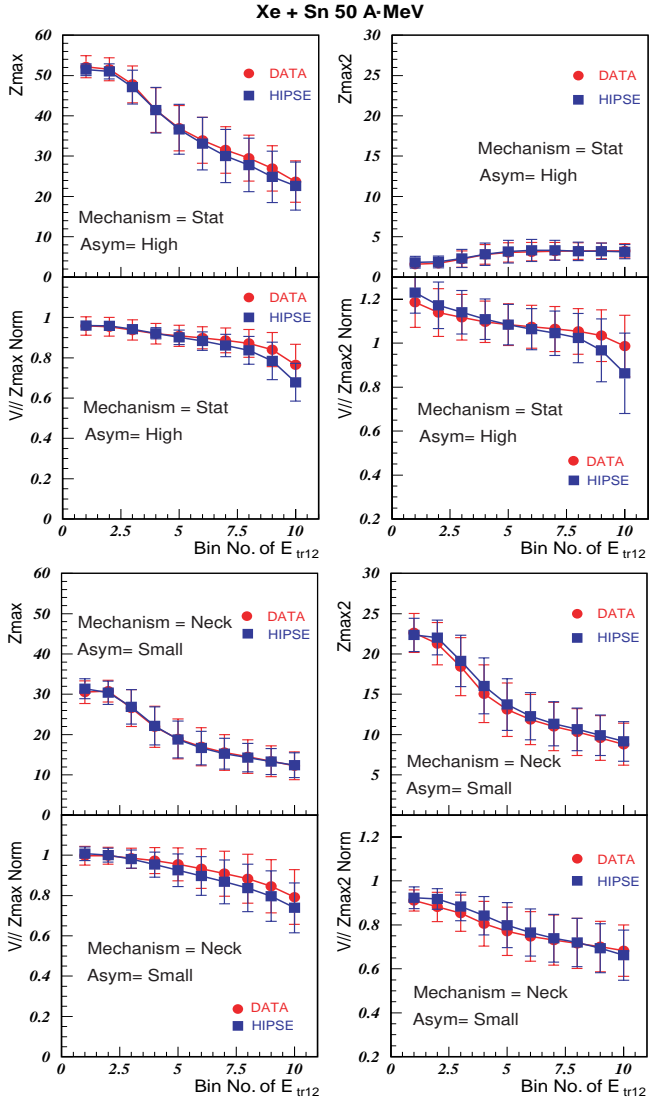}}
\caption{Comparisons of mean charge and the mean parallel velocities of the two heaviest fragments at the front of the center of mas, between HIPSE and the data for various selections of collision violence. This study is made for two types of mechanism during collisions Xe + Sn at 50 A.MeV.}
\label{fig7}
\end{figure}

Figure \ref{fig7} shows, on one hand, the average evolution of the charge of these two fragments according to the bin number of the normalized $E_{tr12}$, and on the other hand, their average parallel velocities in the laboratory frame normalized to the initial projectile velocity, always according to the bin number of the normalized $E_{tr12}$. These two charges are important indicators on the process of de-excitation of the hot nuclei \cite{Borderie1, Frankland1, Pichon1}. The heaviest fragment velocities supply an indication on the fraction of the initial incident energy, dissipated during the reaction. We focus on \textbf{Statistical} collisions with a high asymmetry and collisions with a \textbf{Neck} and a small asymmetry.  In both cases, the agreement between the data and HIPSE is excellent for the reproduction of the average charge of both fragments.

\begin{figure}[htbp]\centerline{\includegraphics[width=8.7cm,height=14.5cm]{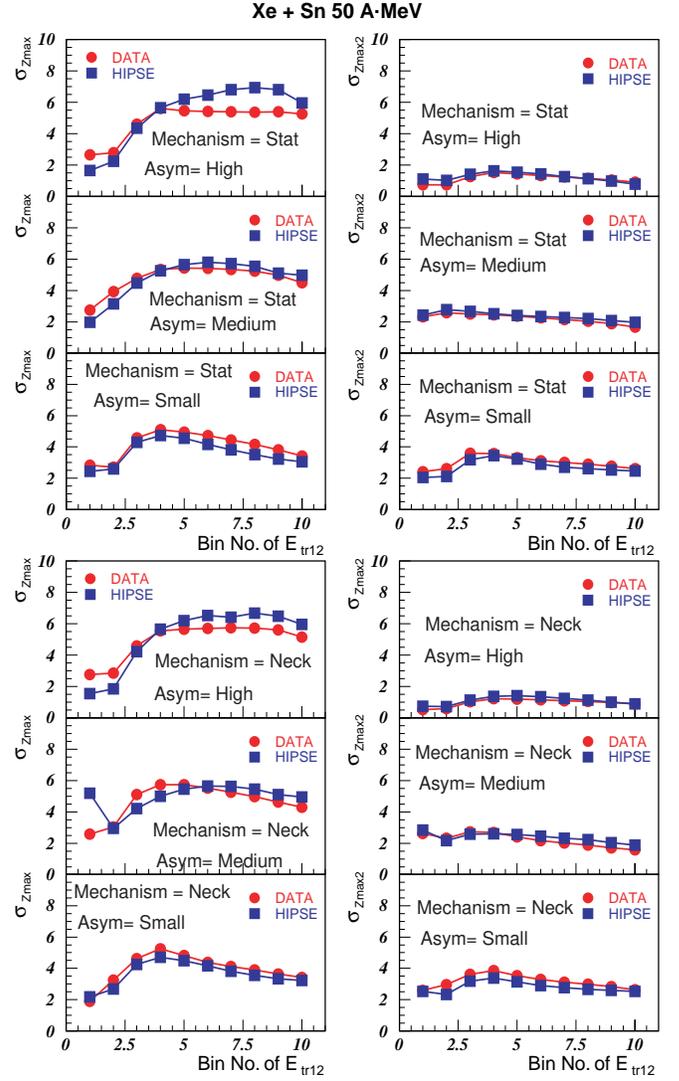}}
\caption{Comparisons of the standard deviations of the distributions of charge of the two heaviest fragments in the front of the center of mass, obtained for HIPSE and the data for various selections of violence of collision. The first column corresponds to the $Z_{max} $ and the second to the $Z_{max2}$. This study is made for both mechanisms and the three studied asymmetries during collisions Xe + Sn at 50 A.MeV.}
\label{fig8}
\end{figure}

 A more global comparison, not presented here, realized for all the reaction mechanisms and for the incident energies 25, 32, 39, 50 and 80 A.MeV, shows a relative discrepancy lower than 10 \% for the heaviest fragment and of the order of 15 \% for the second fragment. The agreement is slightly less good at 80 A.MeV, where, at maximum, differences are of 15 \% and 20 \%. The event generator very reasonably reproduces the dispersion of the data around these mean values as we can see it in figure \ref{fig8}. For all the incident energies, the relative difference between the data and HIPSE concerning the dispersion, is lower, most of the time, than 20 \% for the maximum charge and between 20 \% and 25 \% for the second heaviest according to the asymmetry.
 
In figure \ref{fig7}, we also see that the experimental parallel mean velocity of the biggest fragment deviates from that simulated gradually as the collision becomes more violent. This effect is systematic for all the incident energies and reaction mechanisms. As part of our global comparison, it gives rise to a relative difference of 5\% to 15 \% depending on the violence of the collision for the highest asymmetry for all the mechanisms. There are few \% to 10 \%  remaining for the other asymmetries in the case of a \textbf {Neck} emission and less than 5 \% in the case of \textbf{Statistical} events. For the parallel velocity of second heaviest, the trend is about the same for \textbf {Statistical} events. For \textbf{Neck} events, there seems to be a small dependency on incident energy for small and medium asymmetries. For 50 and 80 A.MeV, the difference is less or about 5 \%. For other incident energies, it can reach 15 \% for medium asymmetry and 10 \% for small asymmetry.

Although HIPSE has shown limitations in reproducing the correct proportions for the various mechanisms, the reasonable agreement, often greater than 10 \%, concerning the physical characteristics of two heaviest fragments at the front of the center of mass and the study of the overall characterictics of the events made in reference \cite{Lacroix1}, encourage us to continue basing our study on this simulation thereafter. We will then see that the physical characteristics of the QP reconstructed by the 3D calorimetry, for data and HIPSE, are very close. In addition, the quality of reproduction of the energetic and angular characteristics of certain LCPs in the QP frame provided by HIPSE \cite{Vient1, Vient2}, validates this choice even more.
 
\section{The study of the ``3D calorimetry''\label{sec3}}
\subsection{Characterization of hot nuclei\label{ssec3.1}}
\begin{figure}[htbp]\centerline{\includegraphics[width=8.7cm,height=15.5cm]{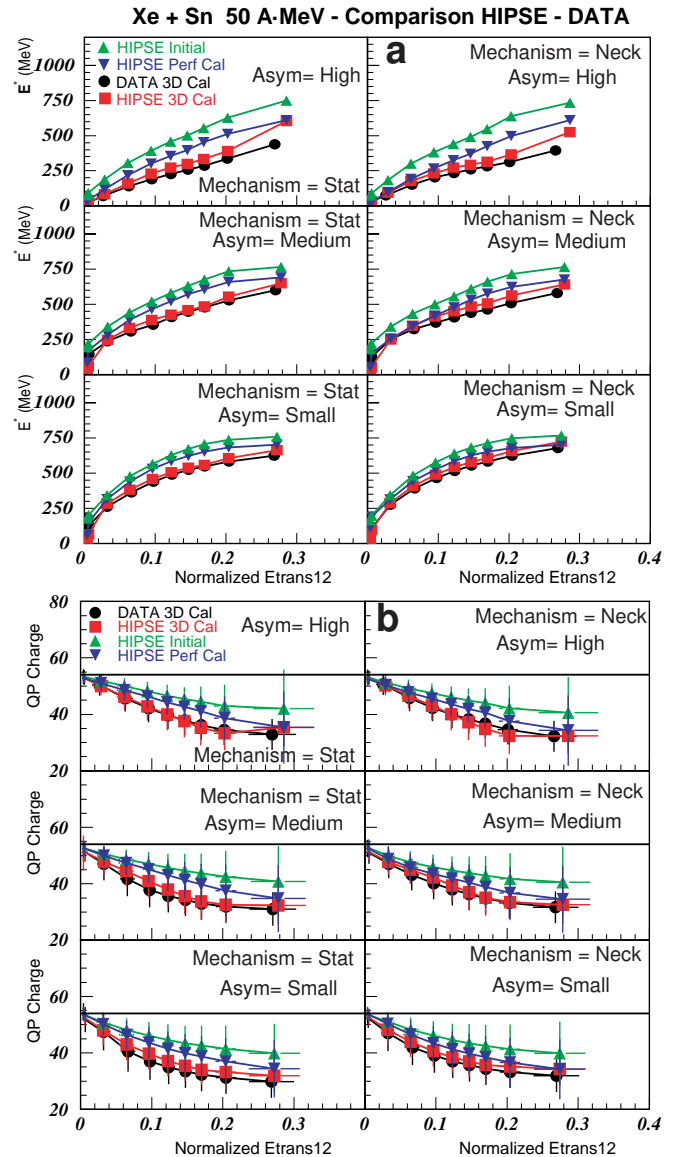}}
\caption{\textbf{a)} Mean correlations between the QP excitation energy and the normalized $E_{tr12}$ found for data using 3D calorimetry and HIPSE using 3D calorimetry, perfect calorimetry and using true initial values.
 \textbf{b)} Mean correlations between the QP charge and the normalized $E_{tr12}$ found for the data by applying 3D calorimetry (full black circles) and for HIPSE by applying 3D calorimetry (full red squares), perfect calorimetry and using the true initial values supplied by the generator. This study is for the two types of mechanism and three asymmetries for Xe + Sn at 50 A.MeV.}
\label{fig9}
\end{figure}

We first chose to study only the system Xe + Sn at 50 A.MeV. Figure \ref{fig9}-a shows the evolution of QP excitation energy and the figure \ref{fig9}-b, the evolution of its charge as a function of the normalized $E_{tr12}$. To deduce these physical quantities, the 3D calorimetry was applied to the real data and HIPSE events filtered by INDRA. To verify the results obtained with 3D calorimetry, we performed a \textbf{perfect calorimetry} authorized by HIPSE that allows us to identify the particle origin in order to select only those emitted by the excited QP.  This is why we applied this \textbf{perfect calorimetry} only with the ions actually detected by INDRA. Since, whatever experimental calorimetry used, it is impossible to do a better calorimetry, we will take this \textbf{perfect calorimetry} as our reference.
We added also the initial mean values of the physical quantities characterizing the QPs generated by HIPSE, which we wish to reproduce.  All these results are presented for the different  selections considered above.

The study of excitation energy clearly shows that we are not able to find the initial excitation energy of the QP, whatever the reaction mechanism and the system asymmetry, even with perfect calorimetry. The effect of the experimental device  therefore seems problematic. The difference between the initial  and reconstructed values tends to decrease with asymmetry. The reaction mechanism  does not seem to play an obvious role. The values obtained with HIPSE seem to be  at 50 A.MeV always slightly higher than those obtained with data.

\begin{figure}[htpb]
\centerline{\includegraphics[width=8cm,height=16.44cm]{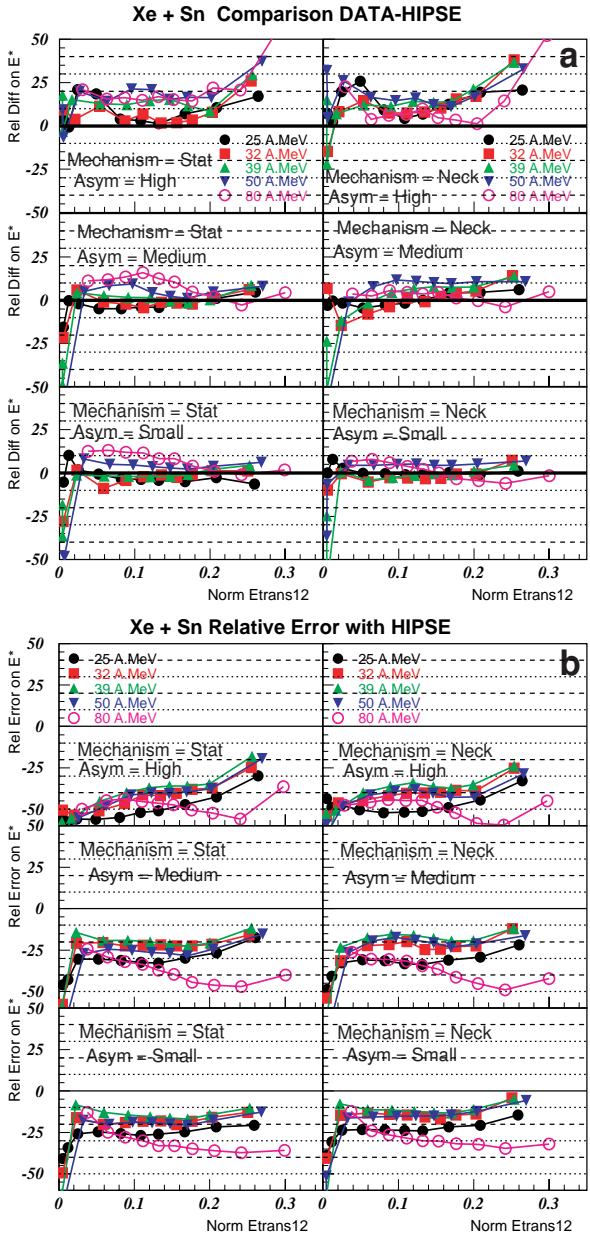}}
\caption{\textbf{a)} Study of the relative differences in \% between the data and HIPSE obtained applying 3D calorimetry to measure the QP excitation energy, for all the studied selections. \textbf{b)} Study of the relative errors in \% on the measurement of the QP excitation energy when 3D calorimetry is applied to HIPSE, for all the studied selections. These studies are made for 25, 32, 39, 50 and 80 A.MeV.}
\label{fig10}
\end{figure}

A more quantitative study of the QP excitation energy is therefore presented in figure \ref{fig10}; on one hand, the relative difference observed between the data and HIPSE, for this physical quantity, when the 3D calorimetry is applied, and on the other hand, the relative error on the measurement, by comparing it with the true initial value, when it is made on the events generated by HIPSE (the calculations of these quantities are explained in the appendix \ref{annexea}). This complementary study is made for all the types of reaction mechanism and asymmetry, at 25, 32, 39, 50 and 80 A.MeV.
It is for the highest asymmetry that the difference  between data and HIPSE is most important, especially for the highest incident energies. Most of time it  remains below 20 \% except for the most violent collisions, where it begins to increase sharply to 40 \% and more depending on the incident energy. When asymmetry decreases, the agreement becomes much better, the difference becomes less than 10 \% and even close to 5 \% for the three lowest incident energies. There is a particular difference for very peripheral collisions linked partially to the poor detection of these collisions in the case of HIPSE but more particularly because of the ``right-left effect'', shown and discussed in the references \cite{Steck2, Vient1, Vient2, Vient3}. It is here more important in HIPSE than in the data because of the kinematics of the model. This tends to increase the apparent energy of the particles in the reconstructed frame.
If we now look at the figure \ref{fig10}-b, we find quantitatively that we are not able to determine the total excitation energy of the QP. With HIPSE, the relative error appears to be independent of the reaction mechanism. On the other hand, it evolves with asymmetry. We underestimate overall by 50 \%, 25 \% and 15 \% when asymmetry decreases. For incident energies, 25 A.MeV and 80 A.MeV, the trends are even a little less good. This may be due to the INDRA filter being too strict for these incident energies, because it is  ``a little less adapted'' to the characteristics of the data supplied by HIPSE, which are more forward focused than the data. 

This complementary study therefore confirms, for all the incident energies, that the 3D calorimetry gives same trends whatever the reaction mechanism. The agreement between HIPSE and the data and the quality of the measure of the excitation energy depends mainly on the asymmetry. 

The study of the charge reconstruction in the figure \ref{fig9}-b is also very interesting. Here again, regardless of the calorimetry used, perfect or 3D, we are not able to determine the initial mean charge of the QP. Perfect calorimetry seems to work a little better than for the excitation energy. The efficiency of detection always plays a major role.  
It is also necessary to note the very good agreement between the data and the simulation, when 3D calorimetry is applied.

\begin{figure}[!h]
\centerline{\includegraphics[width=8cm,height=16.44cm]{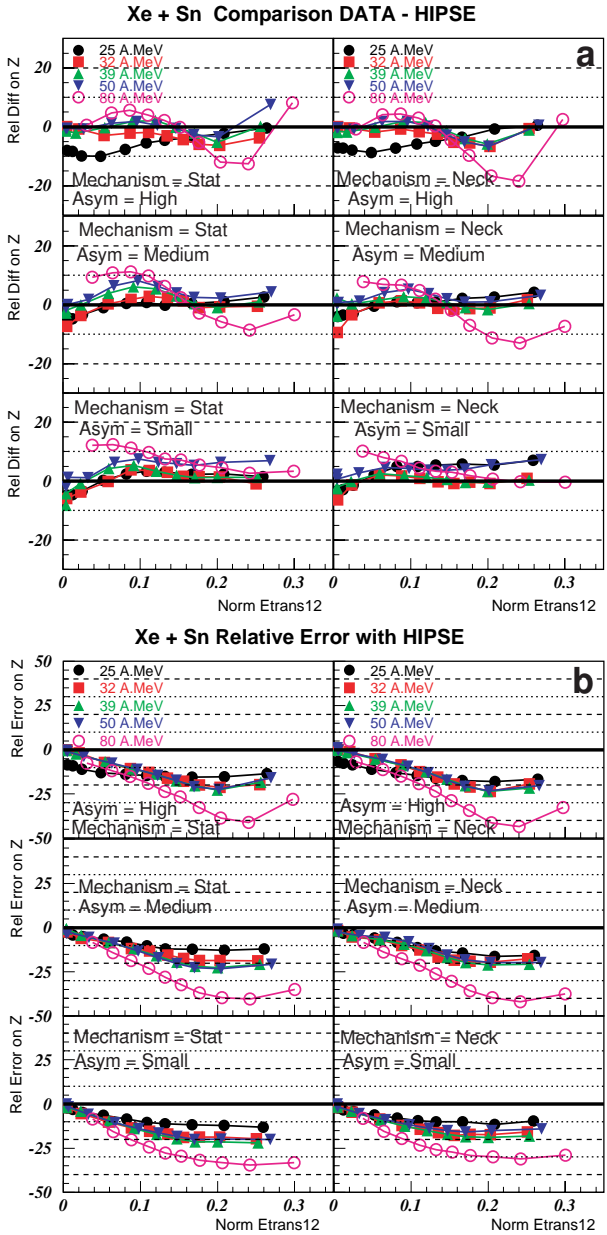}}
\caption{\textbf{a)}  Study of the relative differences in \% between the data and HIPSE obtained applying 3D calorimetry to measure the QP charge, for all the studied selections. \textbf{b)} Study of the relative errors in \% on the measurement of the QP charge when 3D Calorimetry is applied to HIPSE, for all the studied selections. These studies are made for 25, 32, 39, 50 and 80 A.MeV.}
\label{fig11}
\end{figure}

\begin{figure}[htbp]\centerline{\includegraphics[width=8.7cm,height=14.5cm]{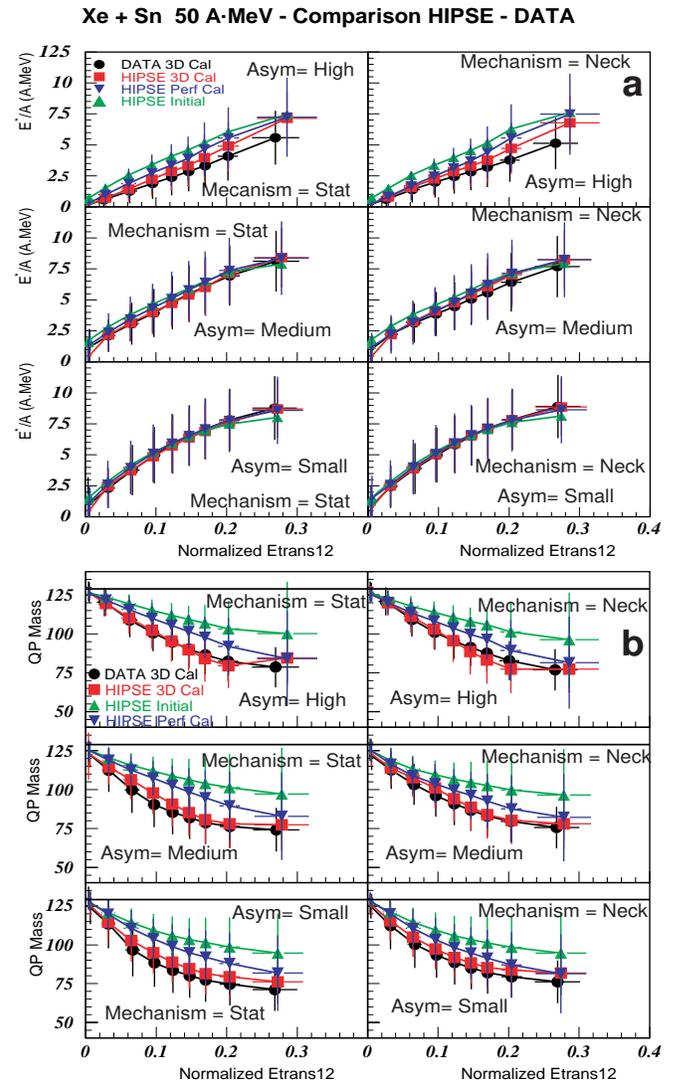}}
\caption{\textbf{a)} Mean correlations between the QP excitation energy per nucleon and the normalized $E_ {t12}$ found for the data applying 3D calorimetry and for HIPSE applying the 3D calorimetry,  perfect calorimetry and using true initial values. 
\textbf{b)} Mean correlations between the QP mass and the normalized $E_ {t12}$ found for the data  applying 3D calorimetry (full black circles) and for HIPSE applying 3D calorimetry (full red squares), perfect calorimetry (full blue triangles) and using true initial values (full green triangles) supplied by the generator. This study is made for the two types of mechanism, three asymmetries during collisions Xe + Sn at 50 A.MeV.}
\label{fig12}
\end{figure}

Figure \ref{fig11} confirms this quantitatively. The relative differences between data and HIPSE are independent of the reaction mechanism. They vary little with asymmetry. Overall, with the exception of the incident energy 80 A.MeV, the difference is less than 10 \% and even 5 \% for semi-peripheral and central collisions whatever the incident energy. As a result, we note that as the reaction violence increases, more and more reaction products coming from QP emission are lost due to detector acceptance. The efficiency of detection decreases with the multiplicity, until saturation. 
Figure \ref{fig11}-b shows that the relative error of charge measurement is only null for the most peripheral collisions then it increases progressively up to about 25 \%. For peripheral collisions, the measurement error changes slightly with the incident energy. Then it becomes more important for 80 A.MeV  when centrality increases, then it reaches  30-40 \%.  Figure \ref{fig12}-b displays the QP mass for the usual selections. All the remarks concerning the QP charge also apply here to the QP mass, since it has been reconstructed from the initial isotopic ratio of the projectile. 

\begin{figure}[htbp]
\centerline{\includegraphics[width=8cm,height=16.44cm]{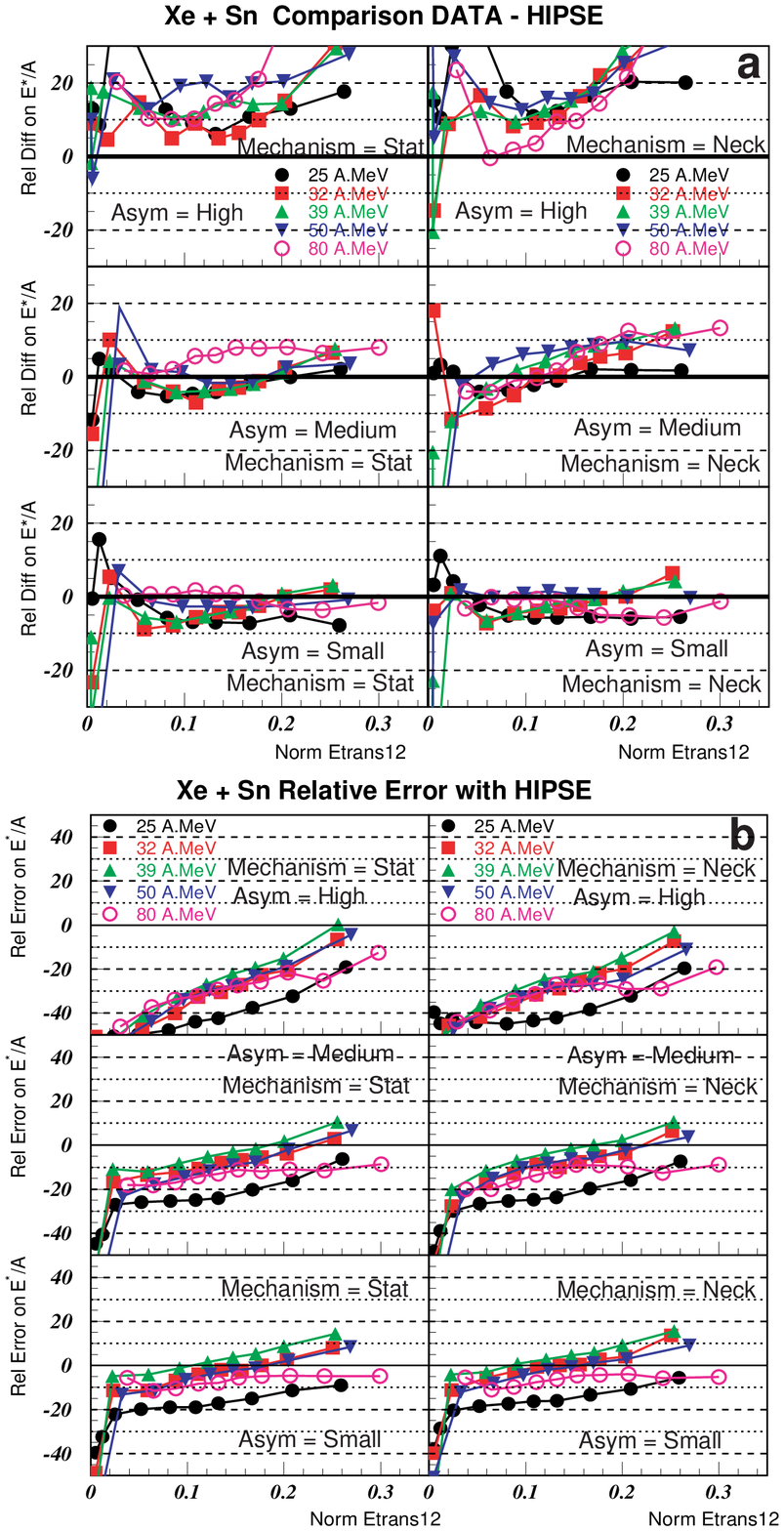}}
\caption{\textbf{a)}  Study of the relative differences in \% between the data and HIPSE obtained applying 3D calorimetry to measure the $E^{\ast}/A$ of the QP, for all the studied selections. \textbf{b)} Study of the relative errors in \% on the measurement of the $E^{\ast}/A$ of the QP when 3D calorimetry is applied to HIPSE, for all the studied selections. These studies are made for 25, 32, 39, 50 and 80 A.MeV.}
\label{fig13}
\end{figure}

In figure \ref{fig12}-a, for medium and small asymmetries, the agreement between data and HIPSE, concerning the measurements of the excitation energy per nucleon, appears quite remarkable considering the observations made previously on mass and excitation energy. It seems that, thanks to compensatory effects, the measurement of this last physical quantity seems better, especially for small asymmetries. In this case, 3D calorimetry applied to the data and HIPSE gives results similar to perfect calorimetry. They are all compatible with the initial value except for very peripheral and central collisions. The quantitative study, presented in figure \ref{fig13}, confirms this fact. There is a reasonable quality of measurement of excitation energy per nucleon for small asymmetries except for very peripheral and central collisions. For high asymmetries,  there is a systematic difference between data and HIPSE for both calorimetries. The excitation energy per nucleon is systematically greater in HIPSE than in the data by nearly 10 \% to 20 \% for almost all incident energies, whatever the studied mechanism. This difference is greatly reduced with asymmetry; since for medium asymmetry, it becomes less than 10 \%. The data become moreover a little higher than HIPSE for the statistical collisions, contrary to those with Neck. For small asymmetries, there are few differences between statistical collisions and collisions with Neck, the data give values larger than HIPSE of 10 \% to 0 \% depending on incident energy and  centrality of the collision. On the other hand, we find again the very poor agreement already observed for very peripheral collisions corresponding to the first three zones of selection according to violence. 
Moreover the error on the measure of $E^{\ast}/A$ follows slightly the same trends. The apparent quality of measurement depends on asymmetry, not on the mechanism of reaction.
For high asymmetries, the excitation energy per nucleon is underestimated of 50 \% for the peripheral collisions. Then, the measure improves with the centrality of the reaction to reach between -10 \% and 0 \%. This trend is the same one for 32, 39 and 50 A.MeV.  
For medium asymmetries, the mean relative errors evolve from -20 \% to 10 \%, while for small asymmetries, it is from -10 \% to 15 \%. The measure for 25 A.MeV is systematically less good of 10 \% approximately. For incident energy 80 A.MeV, the trend appears also a little different, the error on the measure tends rather towards 0 \%.
\begin{figure}[htbp]\centerline{\includegraphics[width=8.7cm,height=14.5cm]{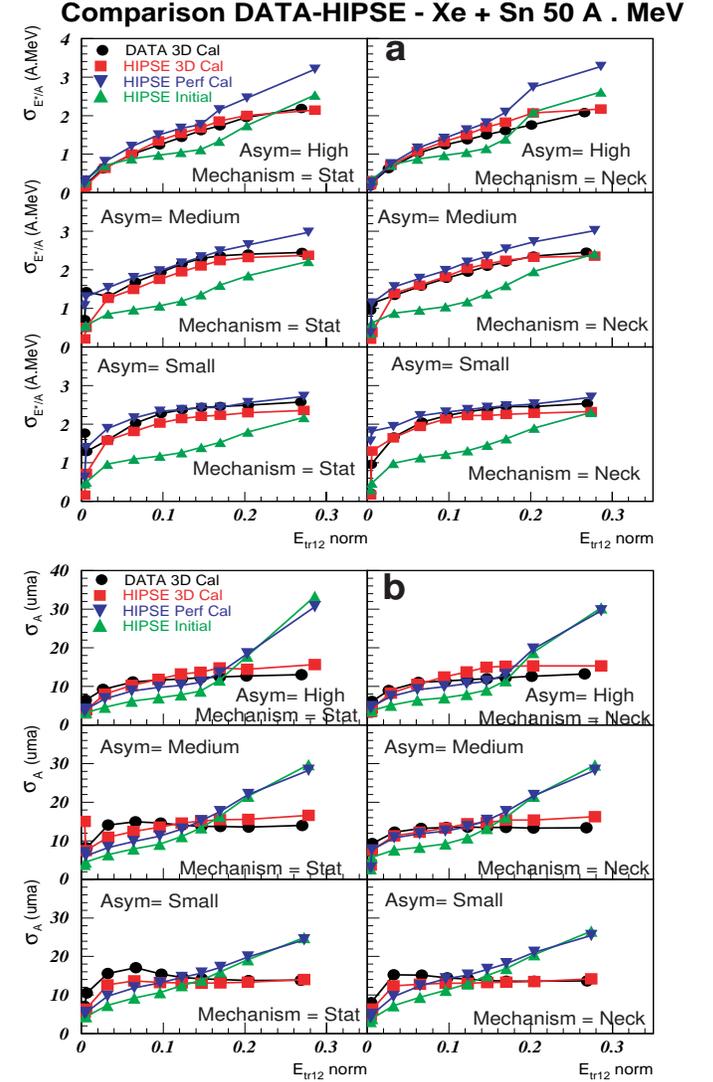}}
\caption{\textbf{a)} Mean correlations between the standard deviation on $E^{\ast}/A$ of the QP and the normalized  $E_{tr12}$ found for the data applying 3D calorimetry  and for HIPSE applying  3D calorimetry, perfect calorimetry and using true initial values. This study is made for both types of mechanism, three asymmetries during collisions Xe + Sn at 50 A.MeV.
\textbf{b)} Mean correlations between the standard deviation on the QP mass and normalized $E_{tr12}$ found for the data applying 3D calorimetry and for HIPSE  applying 3D calorimetry, perfect calorimetry and using true initial values.}
\label{fig14}
\end{figure}

From this analysis, we can therefore draw a certain number of conclusions :

 HIPSE very reasonably reflects the physics of collisions with small or medium asymmetries. 

3D Calorimetry gives results almost independent of the selected reaction mechanism.

The measurement of the excitation energy per nucleon should be taken with caution (see figure \ref{fig13}). We have an accuracy of measurement on this physical quantity which varies according to the asymmetry, the centrality and also to a lesser degree according to the studied system.

There is clearly a crucial influence of the detection device, which acts contradictorily according to the measured physical quantities. The charge and the mass are underestimated due to the efficiency of detection, which seems normal. The measured excitation energy per nucleon can be higher than the original excitation energy per nucleon of the QP, which seems contradictory compared  to the loss of a certain proportion of the particles evaporated by the QP. 

This implies an overestimation of the mean energy contribution of certain particles assigned by our method to the QP, which is also observable in the case of perfect calorimetry.

It also seems important to verify to what extent the calorimetry used increases the width of the distributions of excitation energy per nucleon. Excitation energy per nucleon is a fundamental quantity in nuclear thermodynamics and is often used to sort events and define event classes. We therefore present in figure \ref{fig14}-a for the system Xe + Sn at 50 A.MeV, the standard deviations observed for the various estimations of $E^{\ast}/A$ of the QP and, in the figure \ref{fig14}-b, we added the standard deviations on the mass of the rebuilt QP. Concerning excitation energy per nucleon, the experimental fluctuations, obtained by 3D calorimetry with HIPSE, appear much larger than the initial fluctuations. They seem to increase with  asymmetry without depending on the reaction mechanism, up to about twice as large for small asymmetries. They also do not follow the evolution of true initial fluctuations as a function of normalized transverse kinetic energy. There is a saturation of experimental measurements where the true initial values increase more clearly. The results obtained with the data or HIPSE by applying 3D calorimetry are very close. These are only compatible with a perfect calorimetry for small asymmetries. the perfect calorimetry gives the largest fluctuations, again, because of the influence of detection, and especially the quality of the measurement of particle kinetic energy in the frame of the QP. 
 
 In figure \ref{fig14}-b, fluctuations on the measurement of  QP mass give rise to different trends. For peripheral collisions, the experimental fluctuations, obtained by applying 3D calorimetry, are larger than the true fluctuations. Then, from a normalized transverse energy ranging between 0.15 and 0.17, the latter increase and become larger than the experimental fluctuations, which saturate. Only perfect calorimetry is capable of following the evolution of the true fluctuations by being while always being a little more important, for peripheral and semi-peripheral collisions. Again, the reaction mechanism does not seem to play a role. Experimental fluctuations are similar for data and HIPSE for high asymmetries. They vary for other asymmetries, mainly for peripheral collisions. The data give slightly wider distributions.

\begin{figure}[H]\centerline{\includegraphics[width=8.86cm,height=16.cm]{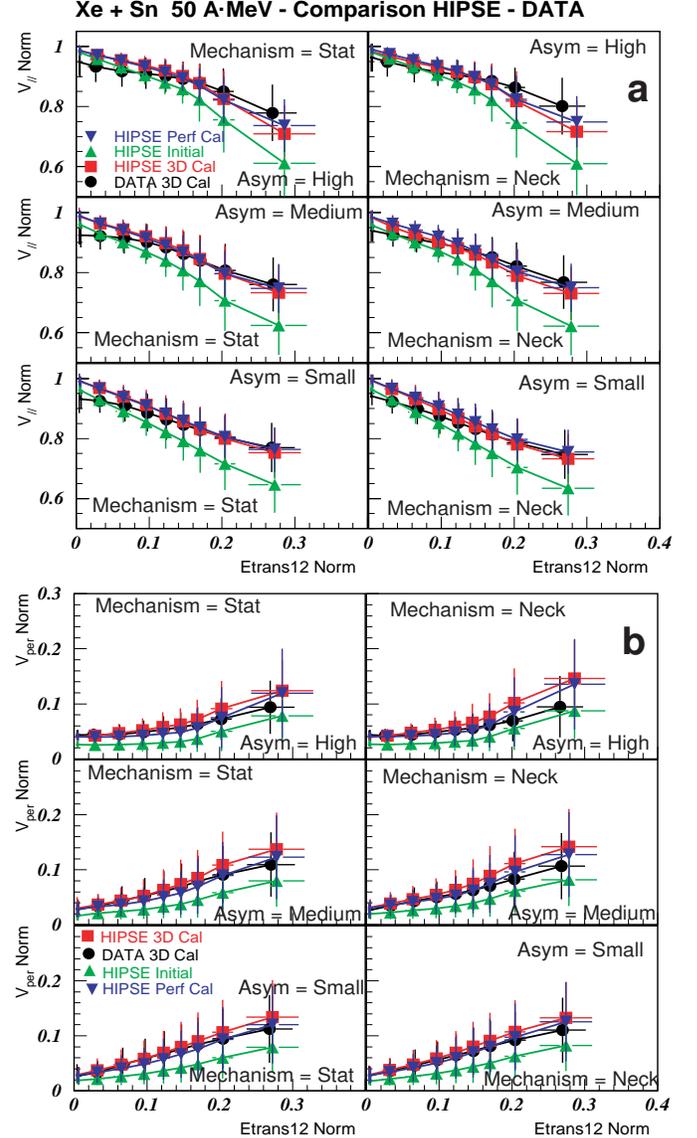}}
\caption{\textbf{a)} Mean correlations between the QP parallel velocity in the frame of the laboratory normalized to the velocity of the projectile and $E_{tr12} $ normalized, found for the data  applying 3D calorimetry and for HIPSE applying 3D calorimetry, perfect calorimetry and using true initial values. This study is made for the two types of mechanism, three asymmetries for collisions Xe + Sn at 50 A.MeV.
\textbf{b)} Mean correlations between the QP perpendicular velocity in the frame of the laboratory normalized to the velocity of the projectile and $E_{tr12} $ normalized, found for the data applying 3D calorimetry and for HIPSE applying 3D calorimetry, perfect calorimetry and using the true initial values.}
\label{fig15}
\end{figure}
The significant result in this figure is the impossibility for our calorimetry  to determine the apparent widening of the QP mass distribution for central collisions. When we do a study similar at 25 A.MeV, this trend is even more obvious. It seems to decrease as the incident energy increases. It should also be noted that it is more present and more important when asymmetry is important.

 This result seems to show that our 3D calorimetry, based on probabilities defined from limited samples of evaporated particles, can only find the average behavior of the measured quantities and  not the fluctuations. An other calorimetry can not do that either  \cite{Vient3}.

\subsection{Study of the hot nucleus velocity\label{ssec3.2}}

We already discussed in the previous sections the fundamental importance of the reference frame to characterize the energetic contribution of evaporation. We will now see to what extent it is possible to return to the kinematics of the QP formed during the reaction, by using our experimental 3D calorimetry. We continue to focus our attention again on the system Xe + Sn at 50 A.MeV. We present in figure \ref{fig15} the parallel and perpendicular components of different velocities, normalized to the velocity of the projectile in the laboratory. We are interested in the QP velocities rebuilt by 3D calorimetry, applied to data and to HIPSE. We also have two references: the initial QP velocity at the moment of the ``freeze-out'' in HIPSE and the velocity rebuild by perfect calorimetry, obtained with HIPSE. 
\begin{figure}[h]
\centerline{\includegraphics[width=8cm,height=16.44cm]{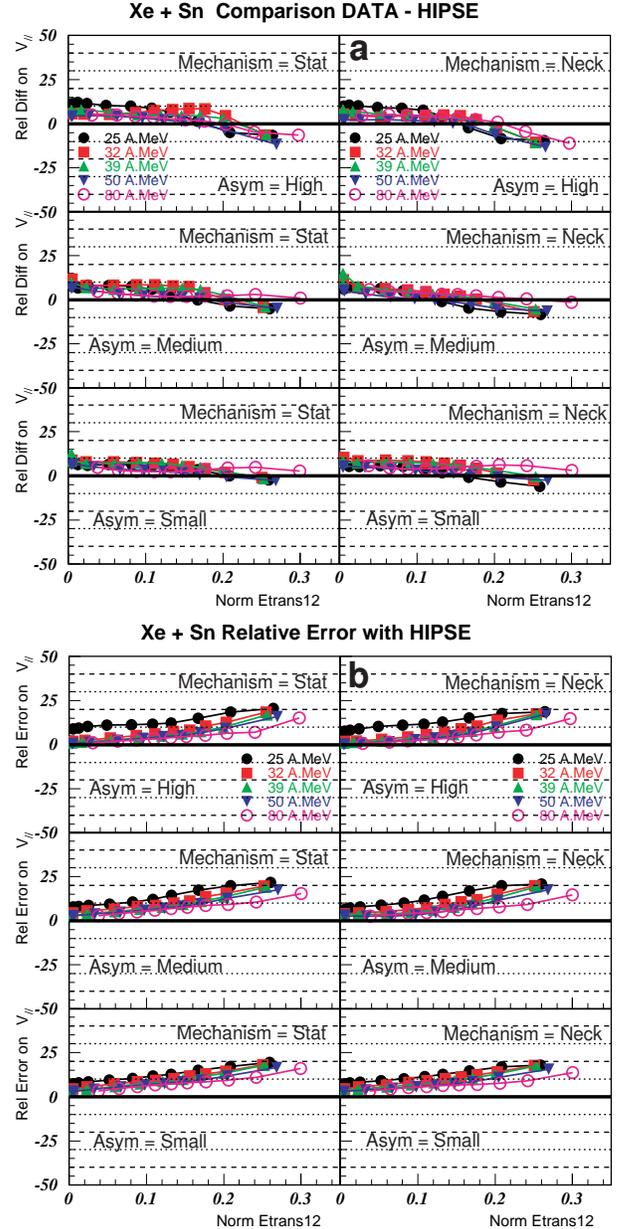}}
\caption{\textbf{a)}  Study of the relative differences in \% between the data and HIPSE obtained applying 3D calorimetry to measure the QP parallel velocity, for all the studied selections. \textbf{b)}  Study of the relative errors in \% on the measurement of the QP parallel velocity when 3D calorimetry is applied to HIPSE, for all the studied selections. These studies are made for 25, 32, 39, 50 and 80 A.MeV.}
\label{fig16}
\end{figure}

The study of the parallel component shows that there is clearly, for all mechanisms and asymmetries, an obvious difference between the initial velocity at the ``freeze-out'' and the different reconstructed velocities. In fact, the opposite would have been abnormal. Indeed, it is necessary to take into account the Coulomb influence of the different partners of the collision during the de-excitation of the QP, which will be accelerated over time by the other participants. This influence is of course greater or lesser depending on the  centrality of the collision, as can be seen in figure \ref{fig15}. Figure \ref{fig16}-b, which gives the relative measurement error on this quantity with HIPSE,  completely confirms this fact for the relative comparison of the measured value and the initial value. The error is independent of the selected mechanism. It is lower for the highest incident energies and decreases with asymmetry, thus with the charge of the heaviest fragment.
 In experiments,  we will never have the opportunity to correct this effect, since we do not have the temporal sequence of emission of the different particles. 
 Figure \ref{fig16}, which  also shows the difference between HIPSE and the data for the parallel component of the reconstructed QP, indicates that the relative variation is not important. It is in the order of 10 \% or less for peripheral collisions and tends towards 0 to -5 \% for central collisions. It depends very few on the reaction mechanism and varies little with incident energy except for 80 A.MeV, which is always a special case and better.

\begin{figure}[h]
\centerline{\includegraphics[width=8cm,height=16.44cm]{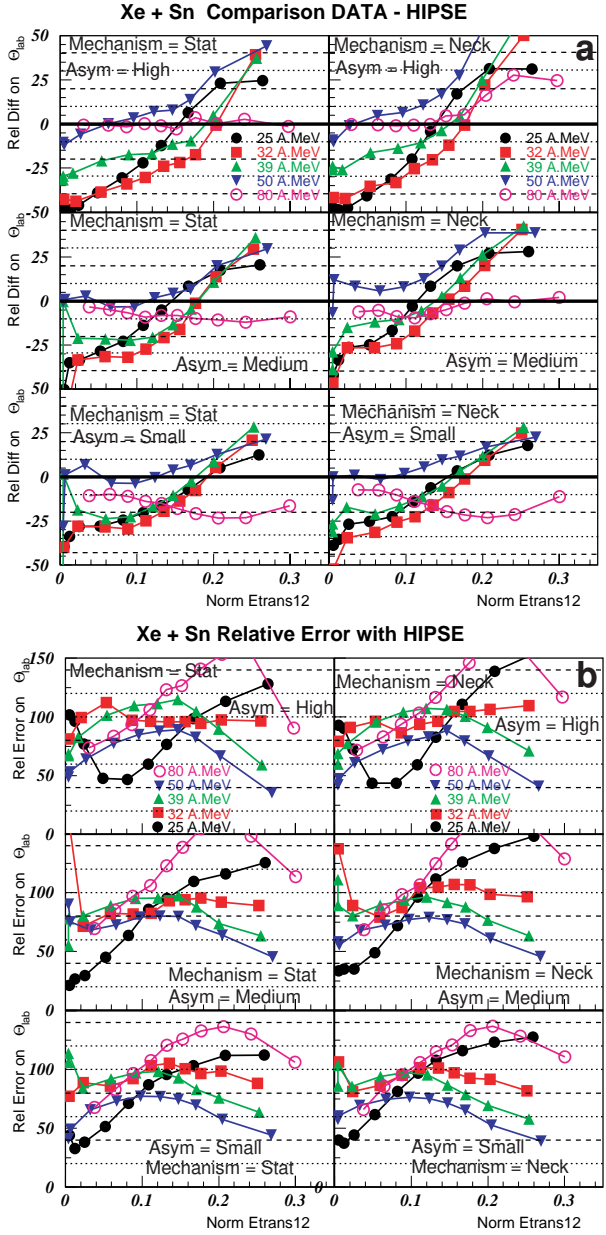}}
\caption{\textbf{a)} Study of the relative differences in \% between the data and HIPSE obtained applying 3D calorimetry to measure the QP polar angle in the laboratory frame, for all the studied selections. \textbf{b)}  Study of the relative errors in \% on the measurement of the QP polar angle in the laboratory frame when 3D calorimetry is applied to HIPSE, for all the studied selections. These studies are made for 25, 32, 39, 50 and 80 A.MeV.}
\label{fig17}
\end{figure} 
It is also important to note the remarkable agreement between 3D calorimetry and perfect calorimetry with respect to the reconstructed parallel component of the QP. 

The data seem a little less compatible with HIPSE. It shows the small differences of kinematics already observed through transverse kinetic energy distributions and the difficulty of  quantitatively reproducing the mechanisms of reaction.

Figure \ref{fig15}-b allows an equivalent analysis of  the component of the reconstructed velocity, perpendicular to the initial beam direction. We find an even greater difference between the initial  and reconstructed values  for all calorimetries. In this case, the Coulomb repulsion must also have an influence. The agreement between 3D calorimetry and perfect calorimetry is apparently less good than for the other component. There is also a larger difference between HIPSE and the data when 3D calorimetry is applied, mainly for the most violent collisions. 

The dynamics of the collision must be partially responsible, as shown in figure \ref{fig17}, where the quality of the measurement of the QP polar angle in the laboratory frame is displayed. First, we observe a clear apparent difference in kinematics between HIPSE and the data whatever the mechanism or asymmetries. We also note a significant relative error on the experimental determination of this angle, between 50 \% and 150 \% in all situations. The Coulomb repulsion between the collision partners must play some role in this trend.

Figure \ref{fig17}-a shows the relative difference between the data and HIPSE, on the mean angle $\theta_{lab}$, the angle between the reconstructed velocity and the initial beam direction  in the laboratory frame. It can be noted that a correct match between HIPSE and the data seems to exist only for peripheral collisions at 50 and 80 A.MeV. It is for these that the role of the re-aggregation is minimal in HIPSE.
The correct treatment of the interaction between the nuclei at small angles must reflect the shape of the edges of the nuclei and must be treated by quantum mechanics.
It is interesting to note that in peripheral collisions, the angular distributions of the QP for 25, 32 and 39 A.MeV  are more focused forward for HIPSE than for the data.

In figure \ref{fig17}-b, we present the error on the measurement of this angle in the HIPSE model, always for the same incident energies. It can be seen that the quality of the measurement depends little on the mechanism and asymmetry, but much more on the incident energy. Whatever the latter, the measure seems bad. The relative error fluctuates between 50 and 150 \%.
This average angle is very sensitive to the experimental device used. The angular resolution limited by INDRA implies a distortion of the angular distribution of particles detected in the frame of the laboratory. We are indeed obliged ``to randomize'' the direction of the velocity of each detected particle over the entire solid angle under which the detection module is seen from the target. This is not done by taking into account the real spatial distribution of particles, because it is not known. Another reason, which could explain this result, is the influence of the beam hole. This prevents the detection of particles at small angles, favoring a larger average angle of the reconstructed QP. Small imperfections which may exist in the software filter can also contribute to the differences between reality and simulation. Finally, we must remember that these angles have mainly a small value, therefore even a small absolute error gives a large relative error.
\section{CONCLUSIONS\label{sec4}}
In a previous article \cite{Vient2}, a new calorimetry of the Quasi-Projectile, aiming to be optimal, was proposed. It tries to take into account in the best possible way the different contributions that can exist in a heavy-ion collision at Fermi energies. This calorimetry is based on the experimental determination of an evaporation probability from the physical characteristics of the particles in a restricted velocity space domain.  

HIPSE  is an event generator trying to integrate all phenomena that can occur at Fermi energies. It is convenient and fast. It seemed interesting to use this generator to study and validate this new calorimetry called ``3D calorimetry''. The application of this experimental calorimetry to the real data and to events generated by HIPSE allowed us to have a some ``control'' of measurements by  constantly comparing them.

The first important  result of this study was the demonstration of the enormous influence of the experimental device, but not as one might expect. The need for the most complete event detection for correct calorimetry generates an extremely spurious effect for some collisions, especially those with a heavy forward-focused evaporation residue, so often with a high asymmetry between the two heaviest fragments at the front of the center of mass.
In fact, the completeness criteria require selecting events with a very particular topology in the velocity space. In fact, in these events, to be detected, the QP residue must necessarily avoid passing through the beam hole made to let the beam pass. The linear momentum conservation requires a LCP contribution sufficiently energetic to permit that. It is the effect, known as ``right-left effect'', already observed \cite{Steck2, Vient1, Vient2, Vient3}, which is largely responsible for the apparent degradation of the experimental QP characterization. In the QP frame, it contributes to an overestimation of the average energy contribution of some particle allocated by calorimetry to the QP. This overestimation is also observed in the case of perfect calorimetry. The non-evaporative contribution, the experimental measurement of the kinetic energy of each particle by INDRA and the determination of the emitter velocity are others sources of errors for 3D calorimetry. These last causes must also be responsible for errors observed for a perfect calorimetry. 

The second important result is that the 3D calorimetry is equivalent to perfect calorimetry when the influence of the ``right-left effect'' is minimal. 

The correction of this effect is complex because it is intimately linked to the collision dynamics, therefore depends on the impact parameter, the incident energy and the studied system. HIPSE reasonably reproduces  the data for small and medium asymmetries, less for high asymmetries. This is also explained partly by this effect, the collision dynamics is not perfectly reproduced, mainly for peripheral collisions.

The fact that the reconstructed velocity obtained by the 3D calorimetry and  the perfect calorimetry are close, whatever the mechanism and the asymmetry, with the except of the last two selections of violence, makes us think that the angular resolution of the detector also plays  a fundamental role in these difficulties of measurement of the excitation energy. We must also bear in mind that, for HIPSE, we compare the kinematics at the ``freeze-out'' moment and a reconstructed kinematics at the end of the cooling. The Coulomb repulsion must also disrupt these measures.

We have not studied the role that the estimation of the neutral contribution can have on the determination of excitation energy. It has yet to be accomplished. It should also be noted that this 3D calorimetry, like other calorimetry, does not allow an accurate estimation of the width of distributions of the physical quantities characteristic of the QP. This disturbs us to make correct  unmixed microcanonical selections of the events. 

For good calorimetry, it is necessary to clearly improve the geometric efficiency, especially  at the front,  granularity, angular resolution and mass resolution of the experimental device. It is a very difficult challenge knowing the experimental qualities of INDRA. 

We have also demonstrated that validation of the new experimental calorimetry, using an event generator and a filter of the experimental set-up, is mandatory. This should also allow to define the parameters of correction for the real data if we want to improve our 3D calorimetry to take into account the defects of the experimental device. But the quality of this correction depends on the realism of these two elements.

\appendix
\section{Principles of calculations of the difference and of the error \label{annexea}} 
To calculate the relative difference on a physical quantity $X$ measured by the experimental calorimetry between the data and HIPSE, we use the following relation:
\begin{equation} 
Rel \: Difference \: on \: X = 100\% \times \frac{X_{HIPSE}-X_{DATA}} {X_{DATA}} 
\label{equa9}
\end{equation} 
To calculate the relative error on a physical quantity $X$ between the result of the measure by the experimental calorimetry and the real initial value $X_{INI}$, for the events supplied by HIPSE and filtered, we use the following relation:
\begin{equation}
 Rel \: Error \:on \: X = 100\%  \times \frac{X_{HIPSE}-X_{INI}} {X_{INI}}
\label{equa10}
\end{equation} 
\newpage 
\bibliography{calorimetryPartII}
\end{document}